%% file: Cr2CoAlArxiv.tex
\documentclass[aps,prl,twocolumn,superscriptaddress,groupedaddress]{revtex4}  
\usepackage{graphicx}  
\usepackage{dcolumn}   
\usepackage{bm}        
\usepackage{amssymb}   
\graphicspath{{Pictures/}} 
\usepackage{epsfig,amsmath,amssymb,color,float,setspace}
\usepackage[utf8]{inputenc}
\usepackage[T1]{fontenc}
\usepackage{titlesec}

\begin{document}

\title{Low-moment ferrimagnetic phase of the Heusler compound Cr$_2$CoAl}
\author{Michelle E. Jamer}
\affiliation{Department of Physics, Northeastern University, Boston, MA 02115 USA}
\author{Luke G. Marshall}
\affiliation{Department of Chemical Engineering, Northeastern University, Boston, MA 02115 USA}
\author{George E. Sterbinsky}
\affiliation{Photon Sciences Directorate, Brookhaven National Laboratory, Upton, NY, 11973 USA}
\author{Laura H. Lewis}
\affiliation{Department of Chemical Engineering, Northeastern University, Boston, MA 02115 USA}
\author{Don Heiman}
\affiliation{Department of Physics, Northeastern University, Boston, MA 02115 USA}

\begin{abstract}
Synthesizing half-metallic fully-compensated ferrimagnets that form in the inverse Heusler phase could lead to superior spintronic devices. These materials would have high spin polarization at room temperature with very little fringing magnetic fields. Previous theoretical studies indicated that Cr$_2$CoAl should form in a stable inverse Heusler lattice due to its low activation energy. Here, stoichiometric Cr$_2$CoAl samples were arc-melted and annealed at varying temperatures, followed by studies of their structural and magnetic properties. High-resolution synchrotron X-ray diffraction revealed a chemically ordered Heusler phase in addition to CoAl and Cr phases. Soft X-ray magnetic circular dichroism revealed that the Cr and Co magnetic moments are antiferromagnetically oriented leading to the observed low magnetic moment in Cr$_2$CoAl.
\end{abstract}

\maketitle

\section{Introduction}

Theoretical calculations have predicated the existence of several inverse Heusler compounds that exhibit zero-moment magnetization while retaining half-metallicity.\cite{Meinerta, SkaftPRB, Galanakisa, Wang2008, Gao2013} These compounds are a subset of SGS materials, where the density of states (DOS, shown in Fig. 1(a)) acts as both a half-metal and a gapless semiconductor.\cite{SkaftAPL, Jamer1, Mn2CoAl1} Such compounds are zero-moment ferrimagnets as their spins are compensated non-symmetrically, which does not prohibit spin-polarization. These compounds are known as half-metallic fully-compensated ferrimagnets (HMFF)\cite{CoeyAPL} and would be attractive for spintronic devices, since their magnetic transition temperatures are often higher than room temperature (400-1000 K). In contrast traditional N\a'eel antiferromagnets cannot be spin-polarized due to their symmetric anti-aligned moments resulting in a symmetric DOS, which was demonstrated in the DOS of the gapless semiconductor V$_3$Al.\cite{CoeyBook,Jamer4} Unfortunately, recent research has shown that some HMFF compounds tend to be unstable and decompose into more stable states, in addition to possessing properties that are adversely affected by structural disorder.\cite{Meinertb, GalanakisJAP, Jamer3}
\begin{figure*}[!tp]
\begin{centering}
	\includegraphics[width = 150mm]{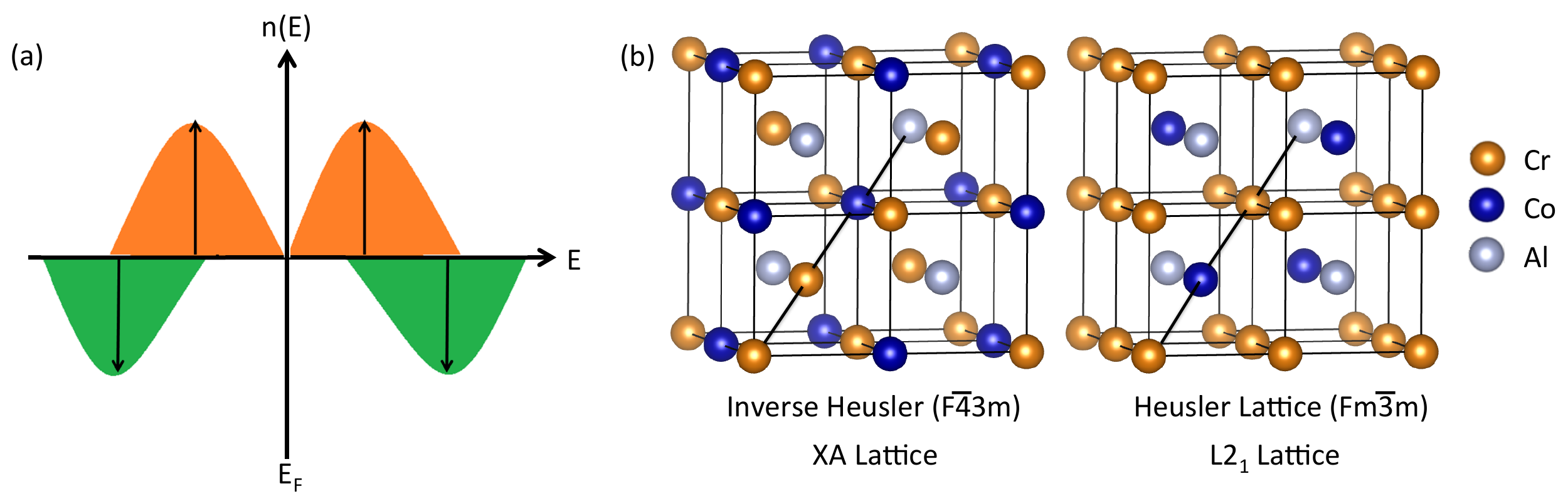}
\caption{\textsf{Schematic illustration of the density of states for the half-metallic semiconductor. The  spin-up (orange) electrons act as a gapless semiconductor whereas the offset in the spin-down (green) carriers allow for half-metallicity. (b) The lattice of both the inverse Heusler (also known as the XA structure) (left) and the L2$_1$ Heusler lattice (right). The atom arrangements along the <111> diagonal are Cr-Cr-Co-Al for the XA structure and Cr-Co-Cr-Al for the L2$_1$ structure.}}
\label{fig:XI1}
\end{centering}
\end{figure*}

This current work focuses on the synthesis and characterization of Cr$_{\text{2}}$CoAl, which has been predicted to be a HMFF when it adopts the inverse Heusler structure. Theoretical calculations have shown that this compound should have a net magnetization of 0.01 $\mu_B$/f.u. and negative energy of formation (-0.27 eV/f.u.), indicating that it is likely to form in the bulk.\cite{Meinertc} The nonsymmetric spin structure of the transition metal atoms (with moments Cr$_{\text{1}}$(1.36 $\mu$$_B$); Cr$_{\text{2}}$(-1.49 $\mu$$_B$); Co(0.30 $\mu_B$))\cite{Meinertc} indicates that this compound can have full spin-polarization. However, calculations have shown that these Cr-based inverse Heusler compounds tend to decompose into other compounds.\cite{Meinertc}

\section{Experimental details}

Stoichiometric polycrystalline Cr$_2$CoAl ingots (1 g) were synthesized using high-purity elements ($\geq$ 99.995 $\%$) via arc melting in an Ar environment. The ingots were annealed at 1000 $^\text{o}$C for 48 hours to homogenize their composition. The samples were cooled to various temperatures between 600 - 1000 $^\text{o}$C for 72 hours to study the effects of annealing on the compound's properties. Using EDS the synthesized ingots' composition varied by < 1 at $\%$ across the samples. Crystallographic properties were probed with high-resolution synchrotron x-ray powder diffraction (XRD) with average wavelength $\lambda$ = 0.459 \AA\ using beamline 11-BM at the Advanced Photon Source at Argonne National Laboratory. The degree of phase segregation was quantified using GSAS for the Rietveld refinement.\cite{Larson, Toby2001, Toby2006} Magnetic properties were determined using a SQUID magnetometer for temperatures between 2 - 400 K. Magnetotransport measurements were measured in the temperature range 2 - 400 K using the van der Pauw method in a refrigerated cryostat placed in the room-temperature bore of a cryogen-free 14 T magnet. XMCD measurements were taken using the total electron yield mode at 300 K using ~70 $\%$ circular polarization of the beam. The measurements were taken at beamline U4B at the Brookhaven National Synchrotron Light Source. The dichroism patterns were taken using both positive and negative fields, and compared well with data using both positive and negative circularly polarized light. XMCD measures the total vector atom-specific moment <m> at the atom's absorption edges.

\section{Structural properties}

The inverse Heusler structure has a space group F\a=43m, as seen in Fig. 1(b) left. The Cr atoms occupy the (0, 0, 0) and ($\frac{1}{4}$, $\frac{1}{4}$, $\frac{1}{4}$) Wyckoff positions. The Co and Al atoms occupy the ($\frac{1}{2}$, $\frac{1}{2}$, $\frac{1}{2}$) and ($\frac{3}{4}$, $\frac{3}{4}$, $\frac{3}{4}$) Wyckoff positions, respectively.\cite{Bansil1999}. The position of these atoms will lead to a 4-atom Cr-Cr-Co-Al basis along the <111> diagonal. This structure is similar to that of the L2$_1$ Heusler phase Fm\a=3m seen in Fig. 1(b) right, which has a 4-atom Cr-Co-Cr-Al basis.\cite{Dubowik2007} Co$_2$CrAl in the L2$_1$ structure was chosen for study due to its large spin polarization and high Curie temperature.\cite{Dubowik2008} Calculations indicate that increased Cr concentration in L2$_1$-structured Co$_{2-x}$Cr$_x$Al drives the magnetization to zero while retaining its half-metallic properties.\cite{Luo2008} Recent work has shown that CoFeCrAl forms in a semi-ordered Heusler lattice with a large atomic magnetic moment.\cite{Kharel} It is challenging to distinguish between the XRD pattern of the XA and the L2$_1$ structure as only a few of the theoretical Bragg peak heights are different. In addition, when atomic mixing occurs, the increased symmetry causes the reduction of the number peaks.\cite{Kulikov1999}

\begin{figure} T$_{\text{C}}$
\begin{centering}
\epsfig{file=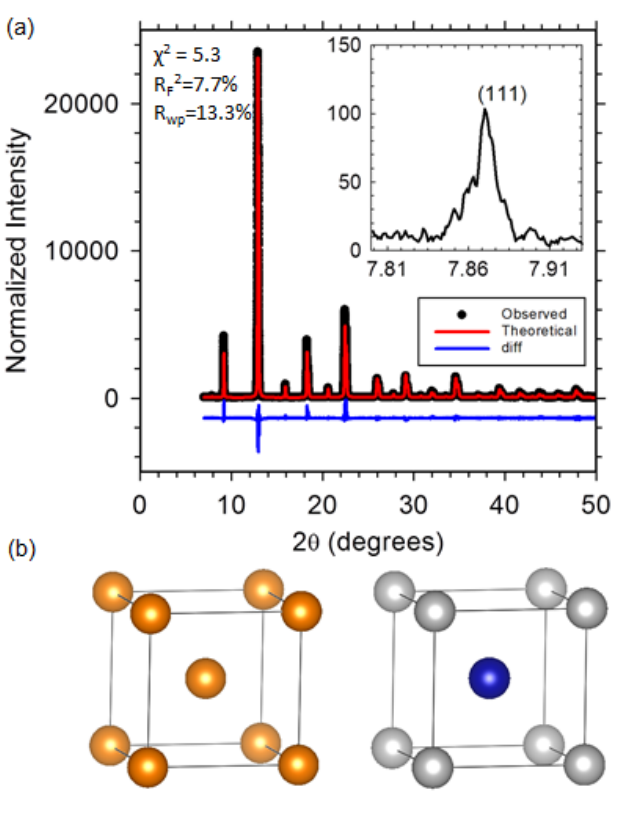, width=70mm}
\caption{(a) The Rietveld refined XRD patterns of the Cr$_2$CoAl sample annealed at 1000 $^\text{o}$C using $\lambda$ = 0.459 \AA. The XRD pattern was refined via GSAS and indicated that there was Cr and CoAl mixed with Cr$_2$CoAl. The fitting parameters are listed in the figure, indicating a good fit. (Inset) The (111) lattice peak is from the Cr$_2$CoAl ordered phase and would not appear if the atoms were highly mixed. (b) The cubic Cr and CoAl bcc lattices.} 
\label{fig:XI2}
\end{centering}
\end{figure}

Figure 2(a) shows the XRD spectrum of a sample annealed at 1000 $^\text{o}$C measured using synchrotron radiation ($\lambda$ = 0.459 \AA) indicating that Cr$_2$CoAl coexists with cubic Cr and CoAl phases (seen in Fig. 2(b)). Previous calculations predicted that Cr and CoAl would form upon the decomposition of Cr$_2$CoAl.\cite{Meinertc} The results of Rietveld refinement (Table 11.1) indicate that the sample consists mainly of disordered Cr$_2$CoAl (48 to 61 vol$\%$) followed by cubic Cr (0.1 to 14 vol$\%$). The only definitive observation of chemically-ordered Cr$_2$CoAl was in the sample annealed at 1000 $^\text{o}$C, which is identified by the <111> peak at 7.9 degrees shown in the inset of Fig. 2(a). The <111> reflection is a necessary condition of the chemically-ordered Heusler phase.\cite{FelserBook}

\begin{table*}
\begin{centering}
\epsfig{file=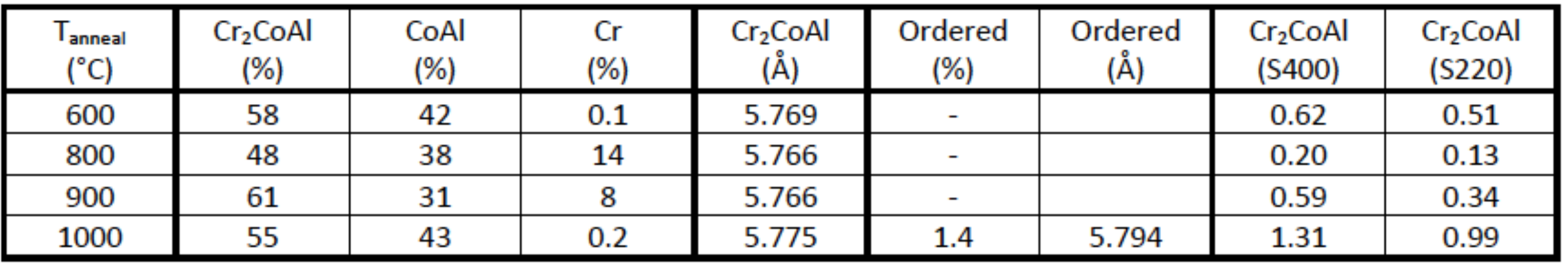,width=\textwidth}
\caption{\textsf{Results of Rietveld refinement of the XRD spectra of Cr$_2$CoAl for various annealing temperatures. The table displays the percentage of each compositional phase and lattice constants. The Cr$_2$CoAl ordered Heusler structure is only achieved for 1000 $^\text{o}$C annealing and has a lattice constant expanded by 0.3 $\%$. The anisotropic strains (S400 and S220) are also shown.}}
\label{tab:XI}
\end{centering}
\end{table*}

The XRD analysis leads to a lattice constant \textit{a} = 5.794 \AA\ for the ordered phase of Cr$_2$CoAl, about 0.3 $\%$ expanded from the disordered phase. This experimental lattice constant matches well with the expected \textit{a} = 5.79 \AA\ for the theoretically-optimized magnetic moment and activation energy for the XA structure.\cite{Meinertc} The Cr lattice parameter varied between \textit{a} = 2.876 - 2.881 \AA, depending on the annealing temperature. The CoAl lattice constant was found to be \textit{a} = 2.868 $\pm$ 0.001 \AA\ and did not vary with annealing temperature. The anisotropic strain S400 and S220 was found for the Cr$_2$CoAl lattice, which broadens the disordered Cr$_2$CoAl peaks. 

\section{Magnetic and transport properties}
SQUID magnetometry was used to measure the magnetic properties of the Cr$_2$CoAl samples as a function of applied magnetic field (H) and temperature. Two magnetic components are identified: (\textit{i}) a paramagnetic (PM) component that varies at the lowest temperatures; and (\textit{ii}) a ferrimagnetic (FiM) component that varies at higher temperatures. It is assumed that these two components are from the two main phase segregates CoAl (PM) and Cr$_2$CoAl (FiM). Fig. 3(a) shows the magnetization M(H) for the chemically-ordered sample annealed at 1000 $^\text{o}$C where it is seen that the PM component increases strongly with decreasing temperature in the range 50 to 10 K. Alternatively, the FiM component (after subtracting the PM component, Fig. 3(a inset) saturates quickly and the saturation moment decreases moderately over a large temperature range, 250 to 400 K. Fig. 3(b) plots the temperature dependence of the two components. The PM component attributed to CoAl rises quickly for T < 50 K. Fitting this component to the Curie-Weiss formula, M = $\frac{C}{T-\theta}$,  where $C$ = 6.7 emu/gK and $\theta$ = 11.5 K. The positive value for $\theta$ can also been found by extrapolating $\chi$ (1/M), confirming the presence of FM interactions. The FiM component attributed to Cr$_2$CoAl (inset,Fig. 3(b)) was determined by subtracting the PM component that was nearly independent of temperature at high temperatures. The data was fit to a mean-field model for T < T$_{\text{C}}$, where M = M$_o$(1-$\frac{T}{T{_{\text{C}}}}){^{\frac{1}{2}}}$, shown by the curve, and the Curie temperature was found to be T${_{\text{C}}}$ $\sim$ 750 K. The large Curie temperature and the low magnetic moment of the ferrimagnetic data indicates that some ordered Cr$_2$CoAl was successfully formed in this sample.
\begin{figure*}
	\centering
	\includegraphics[width = \textwidth]{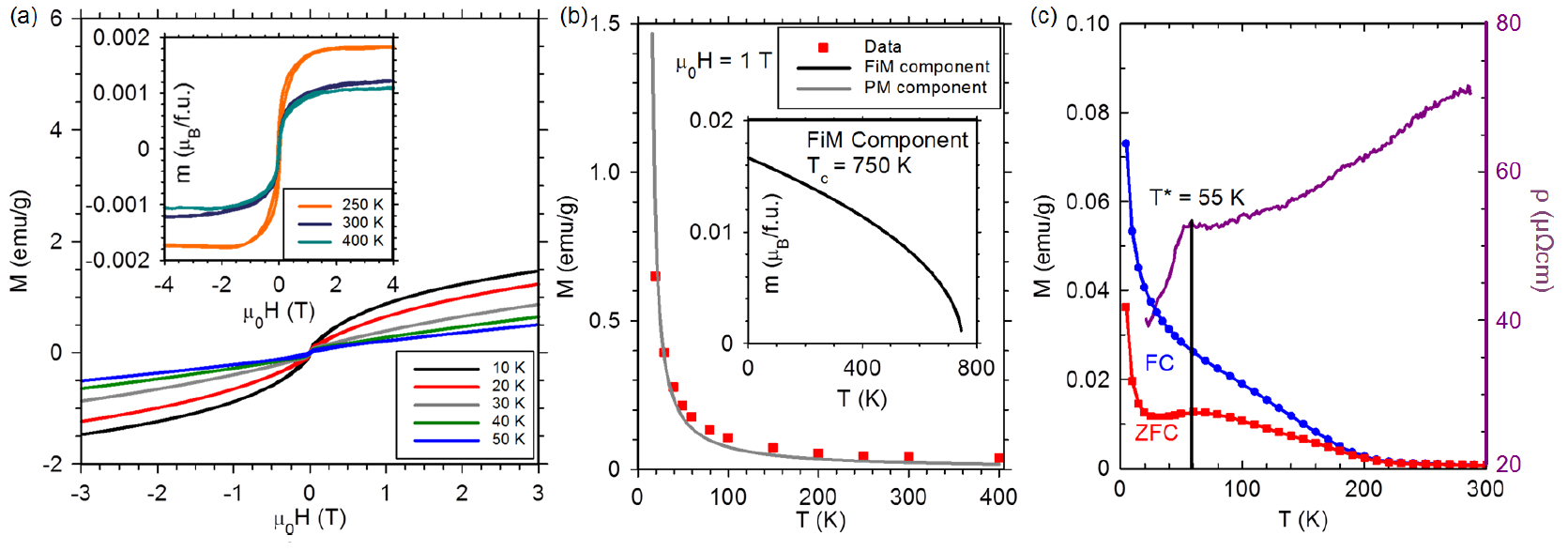}
	\caption{\textsf{Magnetic properties and resistivity of the Cr$_2$CoAl sample annealed at 1000 $^\text{o}$C, which are similar to samples annealed between 600 and 900 $^\text{o}$C. (a) Magnetization versus field for the low temperature and high temperature (inset) components. At low temperatures the dominant PM trend increases below 50 K. In the inset, at high temperatures the FiM component, with PM subtracted, changes moderately (scale in units of $\mu$$_B$/f.u per formula unit). (b) M(T) measured at 1 T, where points are data and curves are model fits. The PM component (grey curve) was fit to Curie-Weiss model M = C/(T-$\theta$), where $\theta$ = 11.5 K. The inset shows the FiM component fit to the mean-field model M = M$_o$(1-T/T$_{\text{C}}$)1/2, where T$_{\text{C}}$ $\sim$ 750 K. (c) Comparison of M(T) measured at 10 mT with the resistivity. The M(T) data shows the spin glass behavior typical of the CoAl FM phase at low temperatures where T*= 55 K. The upper purple curve is the resistivity, which also shows the spin glass transition with a room temperature resistivity of $\rho$ = 73 $\mu$$\Omega$cm.}}
	\label{fig:ds}
\end{figure*}


CsCl-type CoAl can have varying magnetic properties depending on local defects (vacancies and antisites) and small changes in concentration.\cite{Kulikov1999, Vanhatalo1988, Laiho1993} It has been found that Co$_x$Al$_{1-x}$ shows paramagnetic behavior with a negative Curie-Weiss temperature at compositions in the range x = 50.3-50.6 $\%$, and a positive Curie-Weiss temperature for x = 50.9-51.7 $\%$. The Curie-Weiss constant for our PM data is $\theta$ = 12.9 K, indicating that the CoAl phase might be slightly Co-rich and the Co atoms ferromagnetically coupled in CoAl. Fig. 3(c) plots the field-cooled (FC) and zero-field-cooled (ZFC) magnetic data taken at low-field (10 mT). The difference in the FC and ZFC magnetizations indicates that the CoAl phase has a spin glass component at all annealing temperatures. The ZFC curve shows a clear maxima at T* = 55 K, coincident with a spin-glass transition in Co$_x$Al$_{1-x}$, and reported to occur for a slightly Co-rich environment (x = 50.9-51.7 $\%$).\cite{Vanhatalo1988} The low-field ZFC magnetization is compared to the electrical resistivity ($\rho$(T)) in Fig. 3(c) which also shows a cusp at 55 K similar to the spin-glass cusp at T* = 55 K in the ZFC moment. The spin glass transition temperature measured here is consistent with previous measurements of slightly Co-rich CoAl.\cite{Laiho1993, Campbell1982} Previous resistivity measurements on Co-rich Co$_x$Al$_{1-x}$ determined that the composition affects both the resistivity value and temperature minima corresponding to the Kondo effect.\cite{Lee1999, Sellmeyer1972} The overall trend in $\rho$(T) in Fig. 3(c) shows a metallic-like increase for increasing temperature, with a room temperature resistivity of $\rho$ = 73 $\mu\Omega$cm. This value matches with previous data on Co$_x$Al$_{1-x}$ (x = 50.9).\cite{Sellmeyer1972} The resistivity of pure Co$_x$Al$_{1-x}$ follows Matthiessen's rule \cite{Seth1970} where the resistivity is the sum of temperature-dependent and independent parts, but in our case the presence of the Cr and Cr$_2$CoAl phases can also affect the impurity scattering and electron-electron scattering terms.\cite{Lee1999} The effects from the Cr phase are minimal since the metallc Cr is a known spin-density-wave antiferromagnet with T$_N$ = 310 K, which contributes minimally to magnetometry measurements.\cite{Fawcett1988, Ziebeck1982} There is no evidence of magnetic effects from the Cr segregates due to their low concentration and we find no evidence of a N\a'eel transition.

\section{XMCD indicating antiferromagnetically coupled Cr and Co}

\begin{figure*}
\begin{centering}
\epsfig{file=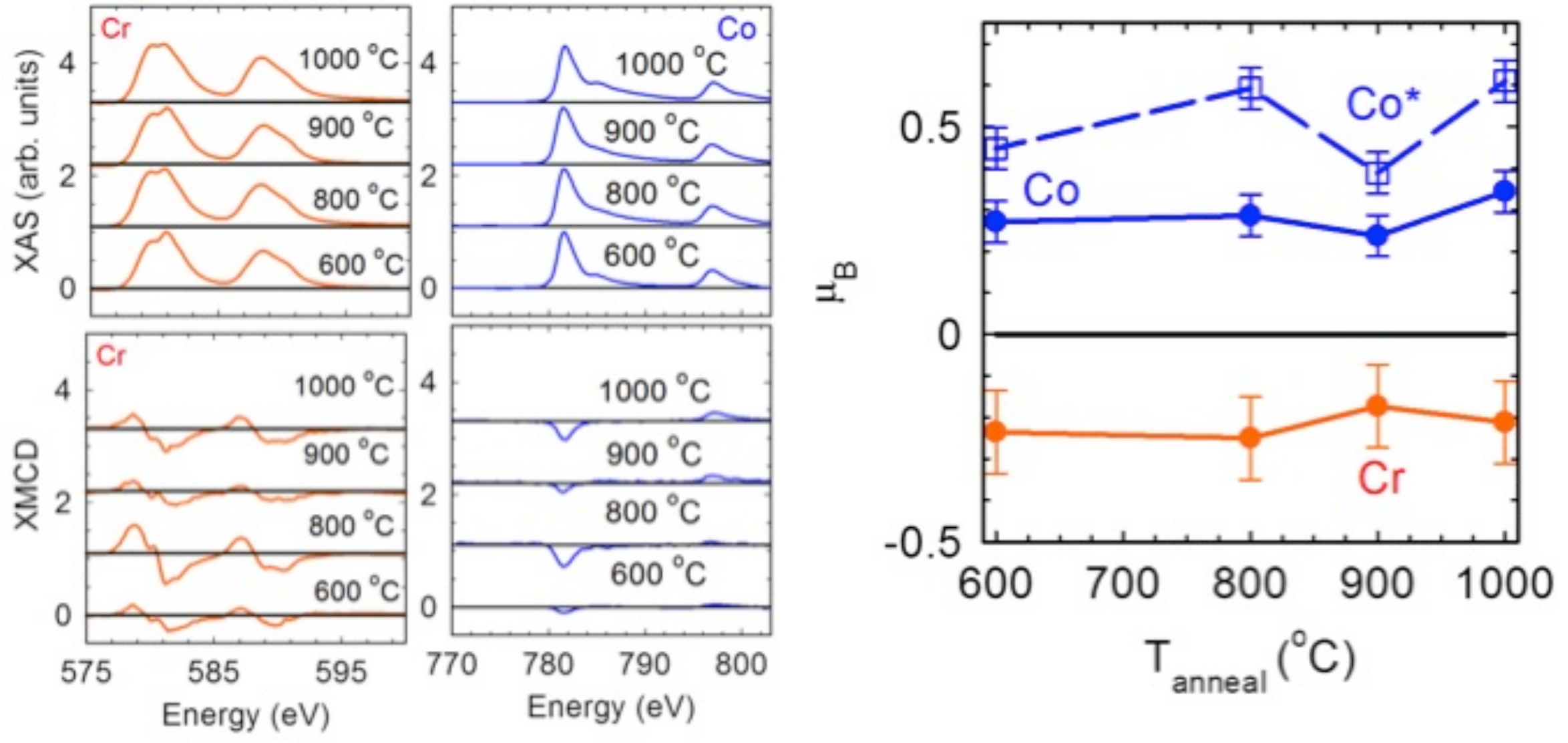, width=140 mm}
\caption{(left upper) The average Cr (left) and Co (right) L-edge XAS patterns for various annealing temperatures. (left lower) The XMCD signal at 1.5 Tesla for the Cr and Co L-edges. (right) The extracted total moment of the Cr and Co atoms. The total moment (labeled "Co*") is the sum of the orbital and spin moments.\cite{FelserBook} The magnetic moment of Co was multiplied by the phase fraction of the Cr$_2$CoAl found through Rietveld refinement, which is depicted by the solid FM-Co line.} 
\label{fig:XI2}
\end{centering}
\end{figure*}

XAS results of the L${_{3}}$ and L${_{2}}$ edges, where the top left and right shows the spectra for Cr and Co shown in Figure 4. The XAS data were first fit to determine the valence states of the Cr and Co \textit{d}-orbitals\cite{Carra1993, Thole1992} and indicate non-integer values that due to the mixed-phase composition. The Cr \textit{d}-orbitals were found to be a mixture of 3\textit{d}$^3$ and 3\textit{d}$^4$ valence states. The effective number of electrons in the \textit{d}-orbitals ranges between 3.2 - 3.6 for the Cr atoms and 5.3 - 6 for the Co atoms.
 
X-ray magnetic circular dichroism (XMCD) spectra for Cr and Co atoms are shown by the lower curves in Figure 4.\cite{Ravel2005} The integrated Cr L$_3$/L$_2$ integrating branching ratios were found to be 1.66 for the ingots, consistent with previous values resulting from strain and small crystallite size determinations.\cite{Wende, LauBranching,Aksov2010} The total extracted atomic magnetic moments are plotted in Fig. 4(b), which is the sum of the orbital and spin magnetic moments.\cite{Jamer3, Laiho1993} The resultant XMCD moment is the average moments of the two Cr atoms in the inverse Heusler lattice. The Cr magnetic moment is small and negative, and within the range of the expected total moment (m$_T$$^{Cr}$ = -0.13 $\mu$$_B$).\cite{Meinertc} (This Cr moment is not anticipated to be affected by the small percentage of antiferromagnetic Cr, and would not contribute to the XMCD signal.\cite{Stohr})  The total Co moment is shown by the square points  and (dashed) solid line in Fig. 4(right), labeled "Co*". However, the measured Co moments are affected by the presence of the paramagnetic CoAl phase in these samples.\cite{Beaurepaire} In order to extract the Co moment associated with the Cr$_2$CoAl phase, the moment associated with the CoAl was subtracted from the total signal by considering the respective phase fractions and assuming that the Co moments have equal contributions to the total Co magnetic signal. This extracted Cr$_2$CoAl moment is represented by the circles and the solid (blue) curve labeled "Co." It is seen that the Cr and Co magnetic moments are antiferromagnetically coupled, with equal but opposite moment, producing a low moment in Cr$_2$CoAl, and agrees with the low total moment observed in the magnetometry results.
 
\section{Summary and outlook}
Cr$_2$CoAl bulk samples were prepared and annealed at various temperatures from 600 to 1000 $^\text{o}$C. Rietveld refinement performed on high-resolution synchrotron XRD indicated that Cr$_2$CoAl was formed at all annealing temperatures. Magnetic measurements indicated the existence of both ferrimagnetic and paramagnetic components contributed by the two main phases in the samples. XMCD measurements allowed extraction of the Cr and Co magnetic moments and confirmed antiferromagnetically coupling. This study indicates that while Cr$_2$CoAl can be synthesized in bulk form, there is a clear need for improved synthesis methods to improve phase purity. In the future, non-equilibrium synthesis of thin films could be expected to provide a better route to single-phase materials that would be required for electronic devices.
\section{Acknowledgements}
\input acknowledgement.tex
\section{References}
\bibliographystyle{unsrtnat} 
\bibliography{cr2coalbib2}
\end{document}

%% file: acknowledgement.tex
We thank T. Hussey for assistance with magnetometry and P. Wei for assistance with preparing samples for transport measurements. We thank D. Arena at NSLS beamline U4B for his guidance. The work was supported by the National Science Foundation grants DMR-0907007 and ECCS-1402738. Use of the National Synchrotron Light Source, Brookhaven National Laboratory, was supported by the U.S. Department of Energy, Office of Science, Office of Basic Energy Sciences, under Contract No. DE-AC02-98CH10886. Use of the Advanced Photon Source at Argonne National Laboratory was supported by the U. S. Department of Energy, Office of Science, Office of Basic Energy Sciences, under Contract No. DE-AC02-06CH11357. We thank the team at 11-BM for the XRD measurements. M.E.J. thanks T. Jamer and K. Jamer for their transportation assistance. M.E.J. is supported by the International Centre for Diffraction Data's Ludo Frevel Scholarship.